\documentclass[aps,prb,reprint,superscriptaddress]{revtex4-1}
\usepackage{epsf}
\usepackage{graphicx}
\usepackage{gensymb}
\usepackage{CJK}
\usepackage{natbib}
\usepackage{amsmath}
\usepackage{upgreek}

\begin{document}

\title{Direct spectroscopic evidence for mixed-valence Tl in the low carrier-density superconductor Pb$_{1-x}$Tl$_x$Te}

\author{P. Walmsley}
\affiliation{Department of Applied Physics and Geballe Laboratory
for Advanced Materials, Stanford University, Stanford, California
94305, USA}

\author{C. Liu}
\affiliation{Ames Laboratory and Department of Physics and
Astronomy, Iowa State University, Ames, Iowa 50011, USA}
\affiliation{Department of Physics, Southern University of Science and Technology (SUSTech), Shenzhen, Guangdong 518055, China}

\author{A. D. Palczewski}
\affiliation{Ames Laboratory and Department of Physics and
Astronomy, Iowa State University, Ames, Iowa 50011, USA}
\affiliation{Jefferson National Accelerator Facility, Newport News, VA 23606, USA}

\author{P. Giraldo-Gallo}
\affiliation{Department of Applied Physics and Geballe Laboratory
for Advanced Materials, Stanford University, Stanford, California
94305, USA}
\affiliation{Department of Physics, Universidad de Los Andes, Bogot\'{a} 111711, Colombia}

\author{C. G. Olson}
\affiliation{Ames Laboratory, Iowa State University, Ames, Iowa
50011, USA}

\author{I. R. Fisher}
\affiliation{Department of Applied Physics and Geballe Laboratory
for Advanced Materials, Stanford University, Stanford, California
94305, USA}

\author{A. Kaminski}
\affiliation{Ames Laboratory and Department of Physics and
Astronomy, Iowa State University, Ames, Iowa 50011, USA}

\date{\today}
\begin{abstract}
Upon doping with Tl the narrow band-gap semiconductor PbTe exhibits anomalously high temperature superconductivity despite a very low carrier density as well as signatures of the Kondo effect despite an absence of magnetic moments. These phenomena have been explained by invoking 2$e$ fluctuations of the valence of the Tl dopants but a direct measurement of the mixed-valency implied by such a mechanism has not been reported to date. In this work we present the unambiguous observation of multiple valences of Tl in Tl-doped PbTe via photo emission spectroscopy measurements. It is shown via our quantitative analysis that the suppression of the carrier density at compositions exhibiting superconductivity and Kondo-like behaviour is fully accounted for by mixed valency, thus arguing strongly against a self-compensation scenario proposed elsewhere for this material and strengthening the case for valence fluctuation models. In addition to the identification of Tl$^{+}$ and Tl$^{3+}$ a possible third intermediate local charge-density is tentatively suggested by full fits to the data, the implications of which are discussed in the context of the charge-Kondo effect.
\end{abstract}


\maketitle

\section{Introduction}
Lead telluride is a narrow band-gap semiconductor that has been influential across a wide range of topics in condensed matter physics and continues to surprise after several decades of study. The present work is focussed on the specific case of thallium-doped lead telluride (Pb$_{1-x}$Tl$_x$Te), that beyond a critical concentration of Tl ($x\approx0.3\%$) exhibits an unconventional superconducting state\cite{Chernik1981, Nemov1998, Matsushita2006}, reaching temperatures of up to $1.5$\,K at hole densities of around 10$^{20}$\,cm$^{-3}$.This is several orders of magnitude greater than any reasonable predictions of the conventional phonon-mediated BCS theory at this carrier density, and the highest of any comparable material by a factor of four \cite{Bustarret2015}. At the same composition that superconductivity emerges a number of experiments have shown evidence that a Kondo-like effect also occurs, but crucially this is not suppressed in magnetic field implying that the effect is not magnetically mediated \cite{Nemov1998, Matsushita2005, Mukuda2018}. The fact that no other hole-dopant produces these effects despite tuning the Fermi energy through the same region of the band structure indicates that a specific property of the Tl impurities must be the origin of these unusual phenomena \cite{Chernik1981, Chernik1981b, Volkov2002, Giraldo-Gallo2016}.

Tl is known to be a valence-skipping element that, owing to the relative instability of its half-filled 6s orbital, has a strong tendency to disproportionate into Tl$^+$ and Tl$^{3+}$ rather than accommodate a 2+ valence. Considering the present case in which Tl is an impurity this implies that the 2+ impurity state may occur at a higher energy than the 1+ impurity state, despite holding fewer electrons on-site (a situation characterised by a `negative effective Hubbard-$U$' \cite{Varma1988}), and this has been supported by calculations \cite{Weiser1981}. In this scenario, the local charge state of the Tl impurity transitions directly from 1+ to 3+ as the chemical potential is lowered through the energy of the Tl 6s orbital (the Tl$^{+}$ impurity state), raising the possibility that the impurity may fluctuate in valence by 2$e$ at the point of degeneracy. Theoretically it has been shown that in Pb$_{1-x}$Tl$_x$Te this negative effective $U$ scenario can provide both a superconducting pairing interaction, and a charge-Kondo effect whereby the valence of the impurity acts as a pseudo-spin by which to map onto the magnetic Kondo problem \cite{Taraphder1991, Dzero2005, Schuttler1989,Hirsch1985,Malshukov1991}.

Existing experimental evidence in support of this picture was recently strengthened by a detailed fermiology study that identifies a resonant impurity state at the Fermi level in superconducting compositions ($x>0.3\%$) of Pb$_{1-x}$Tl$_x$Te that is not present in the non-superconducting analog Pb$_{1-x}$Na$_x$Te at similar energies \cite{Giraldo-Gallo2017, Giraldo-Gallo2016, Nemov1998}. Furthermore, it is argued that the resonant impurity state seems to be enhancing, or even introducing, the pairing interaction rather than simply raising the superconducting critical temperature by increasing the density of states. Whilst it can be argued that these results also imply mixed valency of the Tl impurities \cite{Varma1976}, direct spectroscopic evidence has been lacking to date. Here we report Photo-Emission Spectroscopy (PES) measurements that provide the first direct evidence of multiple valences of Tl in superconducting Pb$_{1-x}$Tl$_x$Te. Two distinct binding energies of the Tl 5$d$ levels are clearly observed in the data, consistent with the Tl impurities taking 1+ and 3+ local valences. The proportion of Tl dopants in each valence state increases as a function of doping in quantitative agreement with the known carrier concentration. The introduction of an unexpected third valence at an intermediate energy improves the fitting of the data, and if intrinsic it indicates that a simple two-level valence fluctuation may not fully capture the physics of the system. These results are a necessary prerequisite for the argument that valence fluctuations are playing a role in the correlated physics in Pb$_{1-x}$Tl$_x$Te.

\section{Methods}

Single crystals of Pb$_{1-x}$Tl$_x$Te were grown by an unseeded physical vapor transport method \cite{Matsushita2007, Matsushita2005}. The overall Tl concentration $x$ was
measured by electron microprobe analysis (EMPA) \cite{Matsushita2005}. The
PES measurements were performed at the PGM beamline
and Ames - Montana beamline of the Synchrotron Radiation Center
(SRC), Stoughton, Wisconsin. All samples were cleaved or scraped \textit{in situ}
along the (100) plane at temperatures between 10\,K and 40\,K.

\begin{figure}
\includegraphics[width=3.3in]{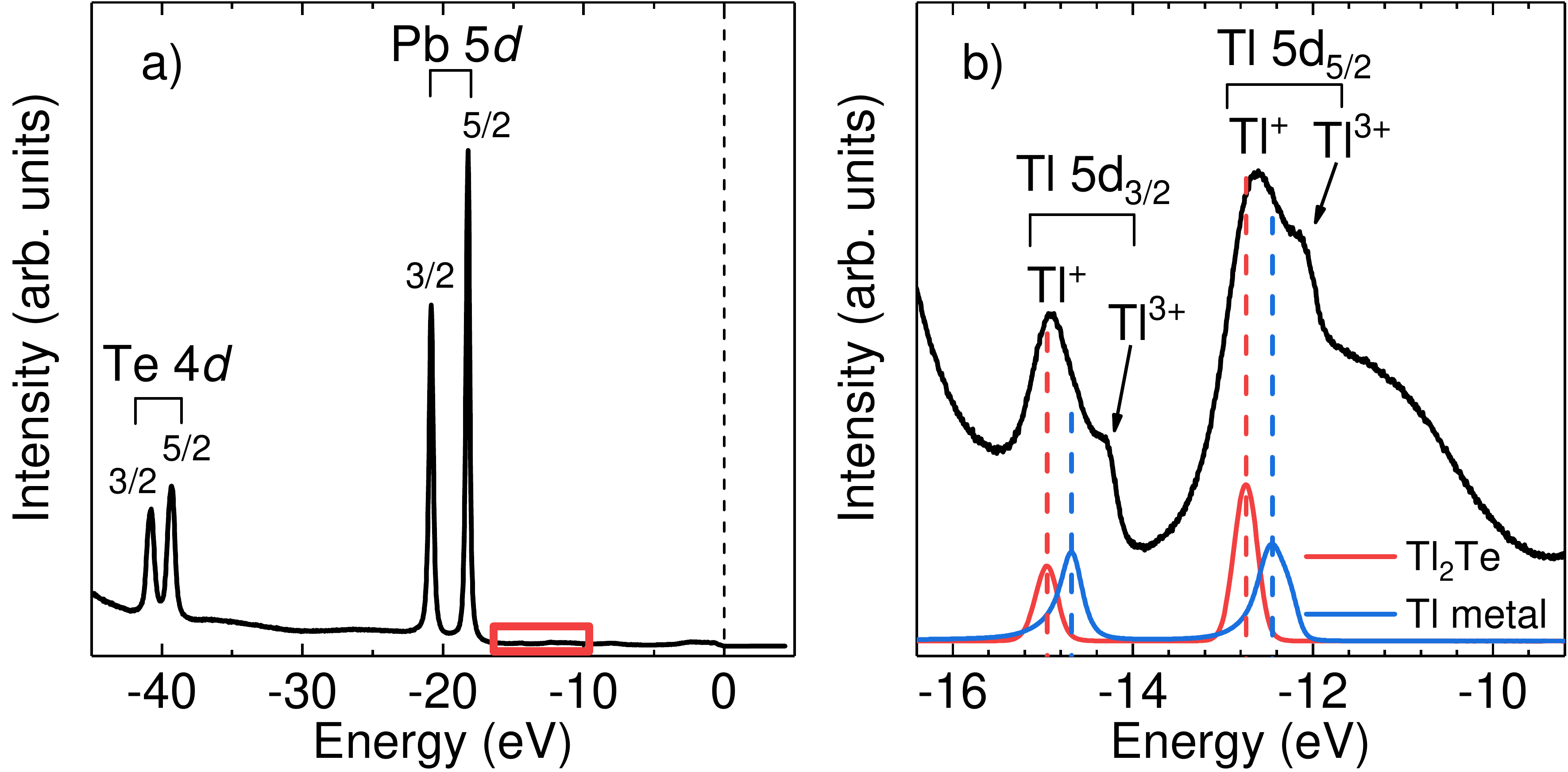}
\caption{Core level analysis of Pb$_{1-x}$Tl$_x$Te by
PES. (a) shows a
typical wide energy range spectra for the core levels of
Pb$_{1-x}$Tl$_x$Te showing the spin-orbit split Te 4$d$ and Pb 5$d$ levels. The more dilute Tl 5$d$ levels are located in the red box, which has been expanded in panel (b). It can be seen in (b) that both the Tl 5$d_{3/2}$ and Tl 5$d_{5/2}$ core levels of Pb$_{1-x}$Tl$_x$Te have multiple peaks. For comparison, the blue curve is a
spectrum of metallic thallium, and the red curve a spectrum of Tl$_2$Te (Tl$^{+}$). These data were taken at temperatures between 10\,K and 40\,K using a Pb$_{1-x}$Tl$_x$Te sample with $x=$1.39\%.} \label{fig1}
\end{figure}

\section{Results}

A representative wide energy range PES spectrum of  Pb$_{1-x}$Tl$_x$Te is shown in Figure \ref{fig1}(a) for a sample with $x=$1.39\%. The  Pb 5$d$ and Te 4$d$ levels are clearly identifiable at their typical binding energies, with the observed splitting due to spin-orbit coupling. The comparatively dilute Tl ions yield a much lower intensity and so are shown in more detail in Figure \ref{fig1}(b). It can be seen clearly in the raw data that both the Tl 5$d_{3/2}$ and Tl 5$d_{5/2}$ levels each show two peaks, which indicates that there are at least two distinct populations of Tl within the measurement volume that are distinguishable by the binding energies of their core electrons. As the binding energy is a function of the local electron density that screens the nucleus (the chemical shift), this  leads to the conclusion that the two distinct populations of Tl must be in different local valence states.

In order to identify the valence associated with the peaks we compare to the Tl 5$d$ peaks in Tl$_2$Te (Tl$^{+}$) as a standard (red line in figure \ref{fig1}(b)). The lower peaks line up with Tl$_2$Te, indicating a 1+ valence that has been observed and inferred previously at $x<x_c$ \cite{Keiber2013, Nemov1998, Matsushita2006, Giraldo-Gallo2017}. Metallic Tl falls approximately equidistant between the two clear peaks, showing that the second peak is not due to inclusions of elemental Tl. The strong tendency of Tl to disproportionate into 1+ and 3+ valences gives a basis on which to assume the second peak is most likely Tl$^{3+}$, and this is strongly supported by previous double-doping studies in which it has been established that Tl becomes a donor when the Fermi energy is tuned below the resonant impurity state (it is an acceptor above it, as it replaces Pb$^{2+}$), and that the total number of states within the resonant impurity states is equal to twice the number of Tl, consistently demonstrating a change of valence by 2$e$. The spectroscopic observation of multiple valences of Tl in PbTe at superconducting concentrations is the key result of this work.

\begin{figure}
\includegraphics[width=3.3in]{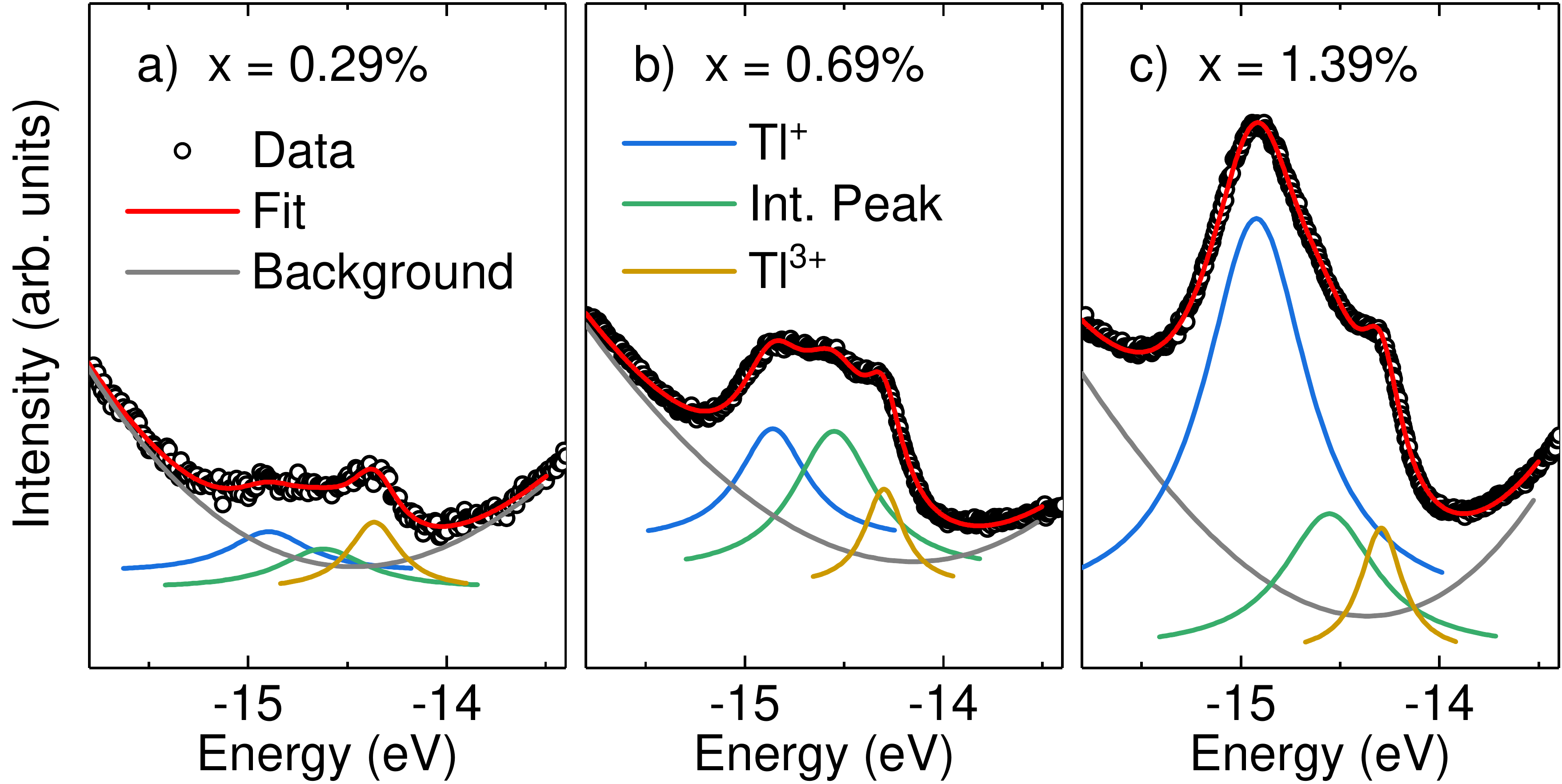}
\caption{Examples of fits to the Tl 5$d_{3/2}$ core levels, shown here for $x=$0.29\%, $x=$0.69\%, and $x=$1.39\% ((a)-(c) respectively). Black circles show the raw data, grey lines are the polynomial background, red curves are the full fitting results, blue (orange) solid lines mark the peak corresponding to Tl$^+$ (Tl$^{3+}$) ions, green lines mark an unexpected third peak that is consistent with the presence of a third population of Tl ions if intrinsic.} \label{fig2}
\end{figure}

To obtain quantitative information, the data were fitted by Lorentzian peaks on top of a smooth polynomial background, with examples shown in Figure \ref{fig2}. This analysis focusses only on the Tl 5$d_{3/2}$ peaks as the shoulder-like peak located at a binding energy of around 11\,eV, formed by the inner band structure of the PbTe lattice, is difficult to fit robustly and would distort results from the Tl 5$d_{5/2}$ levels. Fits to the data were improved by including a third Lorentzian at an intermediate binding energy, close to what might be expected from metallic Tl or an intermediate valence of Tl in PbTe. This is an unexpected result as Tl has a very strong tendency to skip its 2+ configuration in favour of 1+ and 3+, but the presence of inclusions of metallic Tl can also be considered unlikely as discussed later in this report.  It should be noted that, as this intermediate peak isn't clearly visible in the raw data (with the possible exception of $x=$0.69\%), it is possible that its inclusion in the fitting procedure is in fact compensating for an imperfect background subtraction. This could be due to underlying feature at the same energy, an overlap with the tail of the Tl 5$d_{5/2}$ levels (which would give a composition dependence to the fit), or just a poor approximation of the smooth background by the polynomial.

\begin{figure}
\includegraphics[width=3.25in]{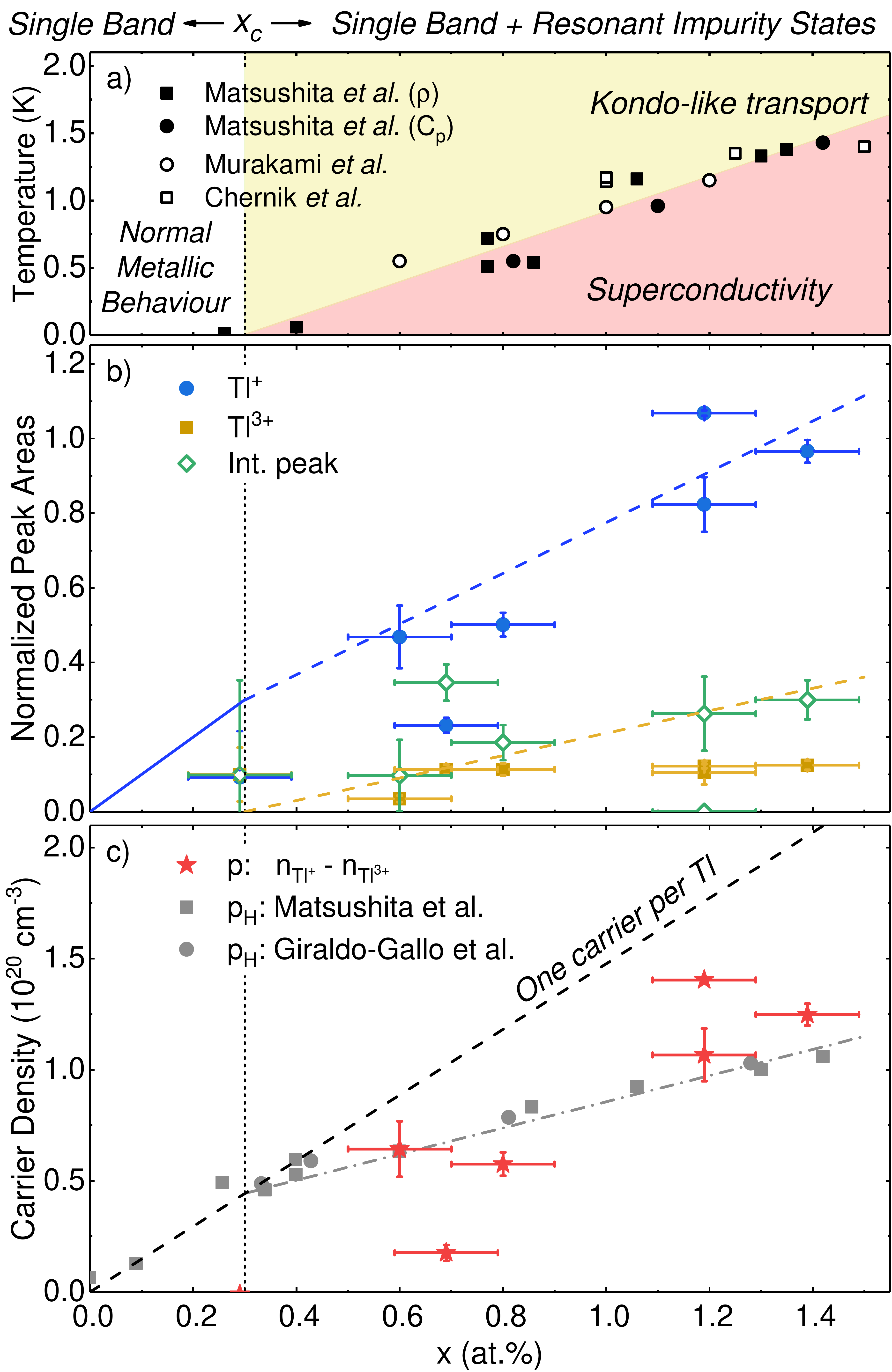}
\caption{(a) A phase diagram illustrating the evolution of the low-temperature electrical transport properties as a function of Tl concentration in Pb$_{1-x}$Tl$_x$Te. At compositions where $x<x_c$ the electrical transport is consistent with a single-band conventional metallic state, but at Tl concentrations above $x_c$ superconductivity, Kondo-like phenomenology, and resonant impurity states at the Fermi level are observed \cite{Matsushita2005,Giraldo-Gallo2017, Nemov1998}. Values of the superconducting critical temperature in single crystals as determined by resistivity and heat capacity (closed black squares and circles respectively) are reproduced from Reference [3], values for thin films and polycrystaline samples (open circles and squares respectively) are reproduced from References [21] and [1]. (b) The evolution as a function of Tl concentration of the integrated areas of the Lorentzian fits, such as those illustrated in Figure \ref{fig2}. The area of the peaks are proportional to the density of Tl dopants in each valence, and normalised such that the total area (total Tl density) matches the measured $x$. The solid blue line indicates the known presence of monovalent Tl$^{+}$ at $x<x_c$, and the dotted blue and orange lines show the expected trends of the Tl$^+$ and Tl$^{3+}$ as implied by the measured carrier density in a two valence model. (c) A comparison between the carrier density as meaured by the Hall effect (gray symbols from References [3] and [17]) and the carrier density implied by the observed peak areas in this study (red stars). The black dashed line indicates the expected behaviour in each Tl contributed a single charge carrier, the gray dot-dashed line is a guide to the eye. In all plots the vertical black dotted line marks the doping at which the superconducting phase and signatures of the Kondo-effect emerge, $x_c=$0.3\%.} \label{fig3}
\end{figure}

Motivated by the phase diagram of Pb$_{1-x}$Tl$_x$Te (reproduced in Figure \ref{fig3}(a)) a quantitative analysis allows us to interpret the observed mixed valency in terms of the suppression of the carrier density that is observed at the same Tl concentrations ($x_c >$ 0.3\%) as superconductivity, Kondo-like transport phenomena, and resonant impurity states at the Fermi level  \cite{Matsushita2006,Murakami1996, Chernik1981,Matsushita2005,Giraldo-Gallo2017, Nemov1998}. The relative proportion of each Tl valence within the measured volume of the sample is estimated by integrating the area of each Lorentzian in the fits described above, the results of which are shown as a function of bulk sample composition (determined by EMPA) in Figure \ref{fig3}(b) (the values have been normalised such that the combined area of the Lorentzians matches the $x$ value). The quantity of each valence increases with $x$ but not at an equal rate as may have been expected in the case of a perfectly sharp resonant impurity state (i.e. perfect degeneracy between two impurity levels that are infinitely narrow in energy). Tl replaces Pb$^{2+}$ in PbTe, making Tl$^+$ an acceptor, Tl$^{2+}$ isovalent, and Tl$^{3+}$ a donor, (Tl$^{0}$ would imply a separate phase and so wouldn't contribute carriers to the bulk phase) and so the behaviour observed in Figure \ref{fig3}(b) where Tl$^{+}$ increases faster than Tl$^{3+}$ implies that the hole concentration should continue to increase in the mixed-valence regime, but at a slower rate than if the system were monovalent. This is indeed observed via measurements of the Hall number\cite{Matsushita2006, Giraldo-Gallo2017}, and we can compare the two datasets by inferring the density of carriers implied by the number density of each Tl valence. The results of this comparison are shown in Figure \ref{fig3}(c) and show good quantitative agreement, albeit with some scatter. Note that the outliers correspond to valences where the Lorentzians corresponding to the Tl$^{+}$ and the intermediate peak are also anomalous in such a way that implies the fits are competing for the same spectral weight, possibly a sign of over-fitting in the data and an indication that the statistical error bars are not representative for these points. These results show very clearly that the Tl$^+$ peak has been correctly identified, particularly as we know that this is the only valence at low compositions from the measured Fermiology (one hole is contributed per dopant at compositions $x<x_c$, as indicated by the solid blue line in Figure \ref{fig3}(b)), because the carrier density continues to increase. They also support the assignment of the Tl$^{3+}$ peak because the compensation at high Tl conentrations gives quantitative agreement with the measured carrier concentration.

\section{Discussion}
These data unambiguously show that there are multiple valences of Tl in Pb$_{1-x}$Tl$_x$Te, in contrast to earlier x-ray and XAFS measurements that lacked the sensitivity to distinguish multiple valences of Tl in such small concentrations and in close proximity to the Pb edge \cite{Waddington1988, Keiber2013}. There has been some debate as to whether a self-compensation model may explain the suppression of the carrier density at $x>x_c$ by retaining a monovalent Tl but introducing Te vacancies \cite{Wiendlocha2018, Kaidanov1985, Nemov1998}, but the present result strongly favours mixed valency as the source of this effect, and this explanation sits more naturally with the concurrent observation of a resonant impurity state. These data do not however differentiate between static and fluctuating mixed valency when taken in isolation. Static mixed valence compounds accommodate mixed valency by allowing distinct bonding environments. For example, in the case of TlS two distinct Tl sites are present and the structure forms a supercell of Tl$^{+}$Tl$^{3+}$S$_2$ \cite{Panich2004}, and in Tl$_2$Nb$_2$O$_{6+\delta}$ the bonding to interstitial oxygen ions is accommodated by a disproportionation of the nearest Tl to give Tl$^{+}_{1-\delta}$Tl$^{3+}_{\delta}$Nb$_2$O$_{6+\delta}$ \cite{Mizoguchi1996}. Whilst chemically interesting these compounds do not exhibit remarkable physical properties as a result of their static mixed valency. The difference in Pb$_{1-x}$Tl$_x$Te is that all of the Tl ions are located on equivalent lattice sites. This may not be intuitive as the orbital occupation is different, but can be understood as the Tl 6$s$ electrons form an inert `lone pair' in PbTe, meaning that they do not significantly contribute to the bonding environment. In the absence of a change in the bonding environment the local fluctuation in the charge density is instead accommodated by the Fermi sea, which allows one Tl to donate electrons via a valence fluctuation provided another accepts electrons via the opposite fluctuation, thus maintaining a constant Fermi level. This interaction between the impurities and the Fermi sea is evidenced by the signatures of the resonant impurity state as discussed in detail in reference \cite{Giraldo-Gallo2017}, but it is also seen in the residual resistivity. If the mixed valency were static, there would be no reason to expect a large difference in the residual resistivity (a proxy for elastic scattering) between the monovalent and mixed valent cases, and any difference would only produce a kink in the residual resistivity as a function of $x$. However, it is seen that for $x>x_c$ the residual resistivity rapidly tends to the unitary limit, implying that there is an additional, strong scattering channel available that is associated with the impurities\cite{Matsushita2007}. 

The results shown in Figure \ref{fig3}(c) provide a basis by which to resolve an unanswered question in the Fermiology of Pb$_{1-x}$Tl$_x$Te, namely why the carrier concentration continues to increase when $x>x_c$ despite the Fermi energy appearing to become quite fixed. As the density of Tl$^{+}$ increases at a faster rate than  Tl$^{3+}$ for $x>x_c$ it implies that the two valences are not strictly degenerate, but are sufficiently close in energy to both be present in the system with a bias towards the lower energy valence,  Tl$^{+}$. This would not be possible for infinitely sharp impurity levels, but resonant impurity states develop a width as they necessarily hybridise with the band structure of the host material provided the orbital characters are not orthogonal \cite{Heremans2012}. This has been observed experimentally in Pb$_{1-x}$Tl$_x$Te via a range of techniques \cite{Nemov1998}. Hence these results imply that while the Fermi level has entered these broadened resonant impurity states, it remains positioned above the mid-point that would give perfect degeneracy between the valences. The width of the impurity state also varies as a function of the density of impurities, and as the maximum density of states is observed to stay constant as a function of Tl concentration the primary effect of adding more Tl must be to increase the width of the resonance \cite{Nemov1998}. As the impurity states widen, a greater portion of them will be found above the Fermi level, thus allowing an increase in the carrier density for a fixed value of the Fermi level.

Finally we must address the origin and implications of the third Lorentzian peak that is used in the fitting of the data. There are three scenarios to consider; the presence of metallic Tl inclusions, an imperfect background subtraction, and the presence of a third Tl valence lying at an intermediate binding energy to the two clearly identified in the raw data. The binding energy of metallic Tl, as shown in Figure \ref{fig1}(b), is very close to the centre of the additional peak, but there are a number of reasons to doubt whether Tl inclusions are present in our Pb$_{1-x}$Tl$_x$Te samples. Most directly, composition maps taken by EMPA with spatial resolution of 1\,$\upmu$m$\times$1\,$\upmu$m could not resolve any inhomogeneity in the samples, which constrains any inclusions to being homogeneously distributed in the sample and well below 1\,$\upmu$m in scale. Secondly, these samples were grown by a physical vapor transport method that doesn't naturally produce phase separation or percolation of impurity phases in the way one might expect from a saturated melt, and even so, we would anticipate that Tl would bond with Te to form Tl$_2$Te during the growth rather than remain in its elemental form (analogously to how NaTe preceipitates in saturated NaTe - PbTe solutions \cite{Yamini2013}). A further argument is that there isn't any clear evidence for a minority superconducting phase from elemental Tl; as Tl has a superconducting T$_c$ of around 2.4\,K, one would expect to see a reduction in the electrical resistivity in the absence of a bulk phase transition in heat capacity, but no such feature is observed.

The motivation for questioning the background subtraction arises as the intermediate peak is not obviously visible in the majority of raw data (with the possible exception of the sample at $x$=0.69\%, as shown in Figure \ref{fig2}b). This is not of course a rigorous argument that the peak does not exist, but it does highlight that the peak is broad and could in principle be approximated by a smooth background, particularly as it is located in the region of the data where the background extrapolation is least constrained. In principle a direct measurement of PbTe in the absence of Tl impurities could clarify this issue, but the background is likely to have a composition dependence as the tails of the Tl 5$d_{5/2}$ and  Pb 5$d_{5/2}$ peaks will change with $x$ and overlap with the region of interest, and so no obvious solution presents itself other than to maintain some slight scepticism regarding the interpretation of this peak.

The most interesting scenario is that there may in fact be three valences of Tl present in the system, with one potential interpretation being that the intermediate peak corresponds to Tl$^{2+}$ as it sits equidistant between the two other peaks and is not required to contribute to the carrier density in order to reconcile the Hall effect data (Tl$^{2+}$ is neutral when replacing Pb$^{2+}$). But despite some support from first-principles electronic structure calculations \cite{Xiong2010} that predict Tl$^{+}$,  Tl$^{2+}$ and Tl$^{3+}$ to have very similar formation energies, this simple version of a three-valence scenario is experimentally ruled out as Tl$^{2+}$ would be magnetic and it has been robustly established that Tl is non-magnetic in Pb$_{1-x}$Tl$_x$Te \cite{Matsushita2005, andronik1988}. It should be considered however that the charge-Kondo effect raises an alternative route to an intermediate local charge density at the Tl impurities. Analogously to how in the magnetic Kondo effect the screening of the magnetic ion produces a Kondo singlet with no magnetic moment at temperatures below the Kondo temperature, in the charge Kondo effect the charge of the impurity is similarly screened by the conduction electrons to give an electrically neutral singlet state \cite{Dzero2005}. The presence of these screening electrons around the Tl impurity leads to a distinct local charge density that must be intermediate to opposingly charged Tl$^+$ and Tl$^{3+}$ in order to be neutral in PbTe, and so consequently this effect could yield an intermediate value of the binding energy of the Tl core levels without invoking magnetic Tl$^{2+}$.

\section{Conclusion}
In conclusion, this work presents the first direct spectroscopic evidence for the mixed-valency of Tl in Pb$_{1-x}$Tl$_x$Te via photo emission spectroscopy. A quantitative analysis of the data shows that the observed mixed valency can fully account for the suppression of the carrier density that occurs at superconducting compositions, arguing against a self-compensation scenario proposed elsewhere. Furthermore, the non-degeneracy of Tl$^+$ and Tl$^{3+}$ observed here leads us to consistently explain the continued increase of carrier density at constant Fermi-energy and Luttinger volume via a broadening of the impurity states, giving a self-consistent understanding of the Fermiology of the system. Full fits to the data were improved by the addition of a third Lorentzian at an energy intermediate to the peaks visible in the raw data. This unexpected additional peak was tentatively attributed to the screening of a fraction of the Tl impurities via the charge-Kondo mechanism, although some caution is required owing to uncertainties in the background subtraction and this tentative assignation requires further investigation.

Mixed valency is an essential ingredient for the valence-fluctuation models that have been invoked to explain anomalous low temperature properties of this material such as superconductivity at low carrier-density and Kondo-like behavior. While this work does not establish a causal link between the observed mixed valency and either of these low temperature electronic properties, it nevertheless definitively establishes that Tl impurities are indeed present as a mixed valence, justifying consideration of such effects, and motivating further investigation of the effects of valence fluctuations in this and related materials.

\section{Acknowledgements}
We thank Yana Matsushita and Ann Erickson for crystal growth. This work was supported by the U.S. Department of Energy (DOE), Office of Science, Basic Energy Sciences, Materials Science and Engineering Division.  Work at Ames Laboratory was supported by the U.S. DOE under contract \# DE-AC02-07CH11358. Work at Stanford was supported by AFOSR Grant No. FA9550-09-1-0583. SRC was supported by NSF DMR 9212658.

\end{document}